\documentclass{ws-ijmpcs}

\usepackage{amsmath,graphicx,verbatim,epsfig}
\usepackage{epstopdf}
\usepackage[utf8x]{inputenc}
\usepackage{amsmath,graphicx}
\usepackage[normalem]{ulem}
\usepackage{datetime}
\usepackage[colorlinks=true,linkcolor=blue,citecolor=blue]{hyperref}
\usepackage{slashed}

\newcommand{\eps}{\varepsilon}

\newcommand{\nn}{\nonumber}

\newcommand{\bn}{{\bar n}}

\newcommand{\veuv}{\varepsilon_{\rm {UV}}}

\newcommand{\be}{\begin{equation}}
\newcommand{\ee}{\end{equation}}
\newcommand{\bea}{\begin{eqnarray}}
\newcommand{\eea}{\end{eqnarray}}
\newcommand{\balign}{\begin{align}}
\newcommand{\ealign}{\end{align}}
\newcommand{\as}{\alpha_s}

\newcommand{\sandwich}[3]{\left< #1 \right | #2 \left | #3 \right >}

\newcommand{\bg}{\begin{gather}}
\newcommand{\foma}{\end{gather}}

\newcommand{\noopsort}[1]{}

\def\ve{\varepsilon}

\def\z{\zeta}

\def\<{\langle}
\def\>{\rangle}

\def\a{\alpha}
\def\b{\beta}
\def\g{\gamma}  \def\G{\Gamma}
\def\d{\delta}  \def\D{\Delta}
\def\l{\lambda}   
\def\s{\sigma}

\def\m{\mu}

\def\z{\zeta}

\def\({\left(}
\def\[{\left[}
\def\){\right)}
\def\]{\right]}

\def\ln{\hbox{ln}}

\def\inf{\infty}

\def \le { \left }
\def \ri { \right}
\def\bp{\bar p}

\newcommand{\ben}{\begin{eqnarray}}
\newcommand{\een}{\end{eqnarray}}

\newcommand{\bef}{\begin{figure}[htb]\centering}
\newcommand{\eef}{\end{figure}}

\def\trimmarks{}

\begin{document}

\markboth{Echevarria, Idilbi, Scimemi}
{On Rapidity Divergences in the Soft and Collinear Limits of QCD}

%
\catchline{}{}{}{}{}
%

\title{On Rapidity Divergences in the Soft and Collinear Limits of QCD}

\author{MIGUEL G. ECHEVARRIA\footnote{
Speaker.}}
\address{Departamento de F\'isica Te\'orica II,
Universidad Complutense de Madrid (UCM),
28040 Madrid, Spain
\\
miguel.gechevarria@fis.ucm.es}

\author{AHMAD IDILBI\footnote{Speaker.}}
\address{European Centre for Theoretical Studies in Nuclear Physics and Related Areas (ECT*),
Villa Tambosi, Strada delle Tabarelle 286, I-38123, Villazzano, Trento, Italy
\\
idilbi@ectstar.eu}

\author{IGNAZIO SCIMEMI}
\address{Departamento de F\'isica Te\'orica II,
Universidad Complutense de Madrid (UCM),
28040 Madrid, Spain
\\
ignazios@fis.ucm.es}

\maketitle


\begin{abstract}
We consider the definition of a transverse-momentum-dependent parton distribution  function (TMDPDF) and its evolution properties given a factorization theorem of Drell-Yan lepton pair production with a non-vanishing transverse momentum. We discuss rapidity divergences that exist in both limits of QCD: the soft and collinear. We argue that only when a specific combination of the soft and collinear matrix elements is formed, rapidity divergences cancel. We also argue that the soft matrix element , calculated in perturbation theory, contains only rapidity divergences and there is no genuine long-distance infra-red divergences.


\end{abstract}


\section{Introduction}	

Factorization theorems (or more appropriately ``factorization statements'') of hadronic high-energy reactions rely on the general idea that the soft and collinear limits of perturbative QCD (pQCD) capture the long-distance contributions of hadronic observables. 
Those observables could either be sensitive or insensitive to the intrinsic transverse momentum of partons inside the colliding hadrons. 
The first case is much more interesting physically as it allows, among other things, to probe the three-dimensional structure of hadrons and also their spin distribution among the hadron constituents.

TMD factorization statements include collinear and soft matrix elements where the dependence on the transverse momentum of partons inside hadrons is manifest. 
When the partonic version of those TMD matrix elements are calculated in pQCD, each one of them contains divergences which are neither ultraviolet (UV) nor long-distance ones. 
Those divergences are referred to as ``rapidity divergences'' (RDs) and they result due to the soft and collinear limits of QCD. 
Such kind of divergences are unphysical in the sense that they cannot be removed by a standard renormalization procedure or by attributing them to a hadronic-scale effects. 
The latter is especially true since in full pQCD (i.e., prior to factorization, or in other words, prior to taking the soft and collinear limits) there are no RDs.
Technically speaking, such divergences appear when Wilson lines, whether soft or collinear, emerge at the level of the soft and collinear matrix elements after the factorization statements have been argued to hold. 

If one adopts the principle that any sensible (``well-defined'') hadronic matrix element should be backed up by a QCD-compatible perturbative calculation, then care must be taken into which kind of divergences are allowed to exist.
According to this principle, one should aspire to cancel all RDs from, encountered in pQCD, all genuine hadronic matrix elements.
This applies to matrix elements with or without polarization dependence as well.

Divergences need to be regularized, however the physical statements should be independent of the specific regulator that is implemented. 
In particular, this should be the case of how a TMDPDF should be defined.

\section{Combining Soft and Collinear: TMDPDF}

For Drell-Yan (DY) TMD spectrum one can write a factorization theorem for the hadronic tensor as follows:
\begin{align}\label{eq:factmodes}
\tilde M &=
H(Q^2)\,
\tilde J^{(0)}_n(\eta_n)\,
\tilde S(\eta_n,\eta_\bn)\,
\tilde J_{\bn}^{(0)}(\eta_\bn)\,,
\end{align}
where $\tilde{J}^{(0)}_{n,\bn}$ stand for the pure collinear matrix elements built from bi-local quark fields separated along the light-cone direction as well as the transverse ones. 
The soft function is the vacuum matrix elements of soft Wilson lines with separation only in the transverse plane. 
Exact definitions of those quantities can be found in~\cite{GarciaEchevarria:2011rb,Echevarria:2012js}. 
$\eta_{n,\bn}$ stand for RDs regulators. 
In general, the need to use \emph{two} different regulators in any regularization scheme results from the fact that two collinear sectors (after factorization) are \emph{decoupled} and the only information that can be transferred between them is through the soft function. 
Among other things, this explains the $\eta$ dependence in Eq.~(\ref{eq:factmodes}). 
The issue of double counting among the soft and collinear modes in Eq.~(\ref{eq:factmodes}) is taken care of since $\tilde{J}^{(0)}_{n,\bn}$ stand for ``pure'' collinear matrix elements, where the soft (or ``zero-bin'') contamination is subtracted out.

Individually, the collinear and soft matrix elements are plagued with RDs. 
This can be seen from an explicit first order calculation in $\as$, where terms of the form $\frac{1}{\veuv}\ln\D$ appear in individual Feynman diagrams and which do not cancel when the proper sum of such diagrams is added up. 
This is unlike the case for the integrated PDF.  
To obtain a well-defined TMDPDF without RDs, a combination of the soft and collinear matrix elements must be formed while also taking into account the decoupling of two collinear sectors. 
This is achieved by noticing that the soft function can be split into two contributions where each contribution has the role of canceling an infinite amount of RDs in each collinear sector. 
This ``infinite'' amount is arbitrary in the same manner as it is the infinite UV divergences to be removed from bare quantities. 
This arbitrariness is the origin of the $\alpha$ dependence in~\cite{Echevarria:2012js} or the $\zeta$ dependence in~\cite{Collins:2011zzd}. 
It should be noted though that such arbitrariness is of no impact when the evolution properties of TMDs are considered.   

Since in full QCD there are no RDs, based on the factorization theorem in Eq.~(\ref{eq:factmodes}) and given that the soft function can be split in a way that incorporates two decoupled collinear sectors, it was shown in~\cite{Echevarria:2012js} that a TMDPDF which is free from RDs can be defined as follows:
\begin{align}\label{eq:tmdnewdef}
\tilde F_{n}(x_n,b;\sqrt{\z_n},\m) &=
\tilde J^{(0)}_{n}(\D^-)\,
\sqrt{\tilde{S}\le(\frac{\D^-}{p^+},\a\frac{\D^-}{\bp^-}\ri)}\,,
\nn\\
\tilde F_{\bn}(x_\bn,b;\sqrt{\z_\bn},\m) &=
\tilde J^{(0)}_{\bn}(\D^+)\,
\sqrt{\tilde{S}\le(\frac{1}{\a}\frac{\D^+}{p^+},\frac{\D^+}{\bp^-}\ri)}
\,.
\end{align}
Here $p^+$ and $\bp^-$ are the large components of incoming quark and anti-quark in DY process. 
$\D^{\pm}$ regulate all divergences which are not UV. 
The $\alpha$ parameter is remnant from the arbitrariness in removing RDs from each TMDPDF, which makes $\tilde{F}_n$ ($\tilde{F}_{\bn}$) dependent on $\z_{n(\bn)}=\alpha Q^2$ ($Q^2/\alpha$). 

In the next section we will argue that the soft function contains only rapidity divergences and there are no infrared (IR) ones.

\section{Soft Function}

\begin{figure}[t]
\begin{center}
\includegraphics[width=0.5\textwidth]{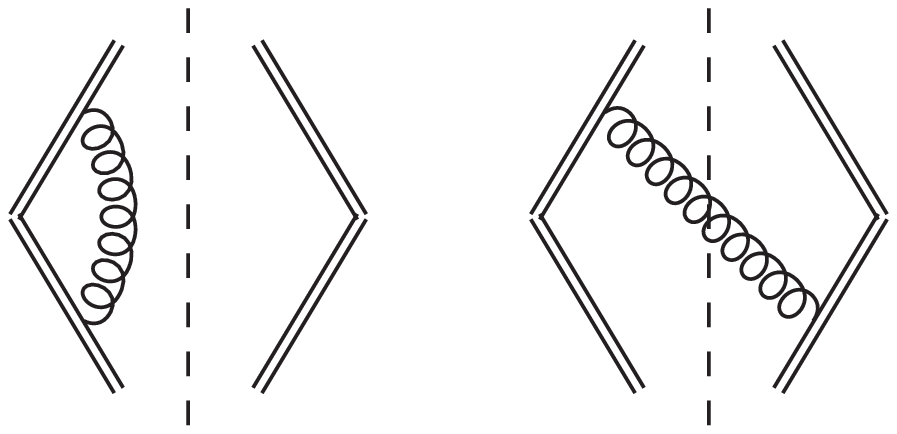}
\\
(a)\hspace{3.0cm}(b)
\end{center}
\caption{\it
One-loop diagrams for the soft function. 
Hermitian conjugate of diagrams (a) and (b) are not shown. 
Double lines stand for soft Wilson lines.
}
\label{fig:soft}
\end{figure}

The relevant soft function for the $q_T$-dependent DY spectrum is given by
\begin{align}
\label{eq:soft}
S(0^+,0^-,\vec y_\perp) =
\sandwich{0}{ {\rm Tr} \; \Big[S_n^{T\dagger} S^T_\bn \Big](0^+,0^-,\vec y_\perp)\le[S^{T\dagger}_\bn S^T_n\ri](0)}{0}\,,
\end{align}
where $S_{n}(x)= P \exp \left[i g \int_{-\infty}^0 ds\, n \cdot A_s (x+s n)\right]$ and the superscripts $T$ stand for soft and transverse Wilson lines, respectively.
The appropriate definitions of the collinear, soft and transverse Wilson lines for DY and DIS kinematics can be found in~\cite{GarciaEchevarria:2011rb}.

Diagram~(\ref{fig:soft}a) and its Hermitian conjugate give the virtual contribution to the soft function,
\begin{align}\label{eq:soft_v}
S_1^{v}&=
-2ig_s^2 C_F \d^{(2)}(\vec k_{s\perp}) \mu^{2 \eps}
\int \frac{d^d k}{(2 \pi)^d} \frac{1}{[k^+-i\d^+] [k^-+i\d^-] [k^2-\l^2+i0]} 
+h.c.
\,.
\end{align}
The poles in $k^+$ are $k^+_1=i\d^+$ and
$k^+_2=(-k_\perp^2+\l^2-i0)/k^-$.
When $k^-<0$ both poles lie in the upper half-plane, so we close the contour through the lower half-plane and the integral is zero.
When $k^->0$ we choose to close the contour through the upper half-plane, picking the pole $k^+_1$, and obtaining (notice that $d^dk=\frac{1}{2}dk^+dk^-d^{d-2}k_\perp$)
\begin{align}
S_1^{v}&=
2 \as C_F \d^{(2)}(\vec k_{s\perp})
\m^{2\ve}\int_{0}^{\infty} dk^- \int \frac{d^{d-2}k_\perp}{(2\pi)^{d-2}}
\frac{1}{(k^-+i\d^-)(k^-i\d^++k_\perp^2-\l^2)}
+h.c.
\,.
\end{align}
Doing now the $k_\perp$ integral we get
\begin{align}
S_1^{v}&=
-\frac{\as C_F}{2\pi} \d^{(2)}(\vec k_{s\perp})
(4\pi\m^2)^{\ve} \G(\ve) \int_{0}^{\infty} dk^-
\frac{(\l^2-k^-i\d^+)^{-\ve}}{(k^-+i\d^-)}
+h.c.
\,.
\end{align}
Finally, performing the integral over $k^-$ and using the $\overline{\rm MS}$ scheme ($\m^2\to\m^2e^{\g_E}/(4\pi)$),
\begin{align}\label{s_virtuals_c}
S_1^{v}&=
\frac{\alpha_s C_F}{2\pi}
\d^{(2)}(\vec k_{s\perp})
\le[\frac{-2}{\veuv^2}+\frac{2}{\veuv}\ln\frac{\d^+\d^-}{\mu^2}
+\ln^2\frac{\l^2}{\m^2} - 2\ln\frac{\l^2}{\m^2}\ln\frac{\d^+\d^-}{\m^2}
+\frac{\pi^2}{6}\ri]
\,.
\end{align}
To get this result we have taken the limits $\d^\pm\to 0$ before $\l^2\to 0$.

The real gluon emission contribution  is given by diagram~(\ref{fig:soft}b) and its Hermitian conjugate,
\begin{align}\label{eq:soft_r}
S_1^r &=
-4\pi g_s^2 C_F \m^{2\ve}\int \frac{d^dk}{(2\pi)^d}
\frac{\d^{(2)}(\vec k_\perp + \vec k_{s\perp})\d(k^2-\l^2)\theta(k^+)}
{(k^++i\d^+)(-k^-+i\d^-)}
+h.c.
\,.
\end{align}
Performing the integral over $k^-$ and $k_\perp$ we get
\begin{align}
S_1^r &=
-\frac{\as C_F}{2\pi} \int_0^\inf dk^+
\frac{1}{(k^++i\d^+)(k_{s\perp}^2-\l^2+i\d^-k^+)}
+h.c.
\nn\\
&=
-\frac{\as C_F}{2\pi}
\frac{1}{|\vec k_{s\perp}|^2+\l^2}
\ln\frac{\d^+\d^-}{|\vec k_{s\perp}|^2+\l^2}
\,.
\end{align}

Using the results above, in Impact Parameter Space (IPS) we have for the virtual contribution
\begin{align}
\tilde{S}_1^v &=
\frac{\alpha_s C_F}{2\pi}
\le[\frac{-2}{\veuv^2}+\frac{2}{\veuv}\ln\frac{\d^+\d^-}{\mu^2}
+\ln^2\frac{\l^2}{\m^2} - 2\ln\frac{\l^2}{\m^2}\ln\frac{\d^+\d^-}{\m^2}
+\frac{\pi^2}{6}\ri]
\,,
\end{align}
while for the real it is
\begin{align}
\tilde{S}_1^r &=
\frac{\alpha_s C_F}{2\pi} \le[
L_\perp^2+2L_\perp\ln\frac{\d^+\d^-}{\m^2}
+2\ln\frac{\l^2}{\m^2}\ln\frac{\d^+\d^-}{\m^2}
-\ln^2\frac{\l^2}{\m^2}
\ri]
\,,
\end{align}
where $L_\perp=\ln(\m^2b^2e^{2\g_E}/4)$.
We have used the following identity in $d=2-2\ve$ to perform the Fourier transforms:
\begin{align}
\int d^d\vec k_\perp e^{i\vec k_\perp\cdot \vec b_\perp}
f(|\vec k_\perp|)
&=
|\vec b_\perp|^{-d} (2\pi)^\frac{d}{2} \int_0^\infty dy\, y^\frac{d}{2} J_{\frac{d}{2}-1}(y)\,
f\left(\frac{y}{|\vec b_\perp|}\right)
\,,
\end{align}
with the particular results
\begin{align}
\int d^d\vec k_\perp e^{i\vec k_\perp\cdot \vec b_\perp}
\frac{1}{|\vec k_{\perp}|^2+\l^2}
&=
\pi\, \ln\frac{4e^{-2\g_E}}{\l^2 b^2}\,,
\nn\\
\int d^d\vec k_\perp e^{i\vec k_\perp\cdot \vec b_\perp}
\frac{\ln\le(|\vec k_\perp|^2+\l^2 \ri)}{\le(|\vec k_\perp|^2+\l^2 \ri)}
&=
-\pi K_0(b\l)\, \ln\frac{b^2}{4e^{-2\g_E}\l^2}
\,,
\end{align}
that can be obtained by setting $d=2$ right from the start because there are no UV divergencies, and the IR ones are regulated by $\l^2$.
We have also used the following expansion for the Bessel function
\begin{align}
K_0(b\l) &= \frac{1}{2}\ln\frac{4e^{-2\g_E}}{b^2\l^2}
+{\cal O}\le((b\l)^2\ri)
\,.
\end{align}

Finally, combining virtual and real contributions in IPS, we obtain the soft function at ${\cal O}(\as)$,
\begin{align}\label{eq:soft_res}
\tilde S_1 &=
\frac{\alpha_s C_F}{2\pi}
\le[\frac{-2}{\veuv^2}+\frac{2}{\veuv}\ln\frac{\d^+\d^-}{\mu^2}
+L_\perp^2+2L_\perp\ln\frac{\d^+\d^-}{\m^2}
+\frac{\pi^2}{6}\ri]
\,.
\end{align}
Thus we see that all the dependence on $\l^2$ cancels and there is just the $\delta^{\pm}$ which regularizes only RDs, inherent to the introduction of  (soft) Wilson lines in the soft function. 
As mentioned before, those RDs will cancel the ones in the pure collinear matrix elements and the only remaining divergence in each TMDPDF will be just the collinear IR divergence. 
Thus eventually one is able to perform an OPE of the TMDPDF onto the integrated PDF for intermediate values of the transverse momentum.

\section{Evolution of TMDs}

The evolution of TMDs is given in IPS as
\begin{align}\label{eq:tmdevolution}
\tilde F(x,b;Q_f,\m_f)& = \tilde F(x,b;Q_i,\m_i)\, \tilde R(b;Q_i,\m_i,Q_f,\m_f)\,,
\end{align}
where the evolution kernel $\tilde R$ is~\cite{GarciaEchevarria:2011rb,Echevarria:2012js}
\begin{align}\label{eq:tmdkernel}
\tilde R(b;Q_i,\m_i,Q_f,\m_f) &=
\exp\le\{
\int_{\m_i}^{\m_f} \frac{d\bar\m}{\bar\m} \g_F\le(\as(\bar\m),\ln\frac{Q_f^2}{\bar\m^2} \ri)
\ri\}
\le( \frac{Q_f^2}{Q_i^2} \ri)^{-D\le(b;\m_i\ri)}
\,.
\end{align}
By applying the renormalization group invariance to the hadronic tensor $\tilde M$ in Eq.~(\ref{eq:factmodes}) we get
\begin{align}\label{eq:drg}
\frac{dD}{d\ln\mu}=\G_{\rm cusp}
\,,
\end{align}
where $\G_{\rm cusp}$ is the cusp anomalous dimension in the fundamental representation, known up to 3-loops~\cite{Moch:2004pa}.
Matching the perturbative expansions of the $D$ term,
$D(b;\m) = \sum_{n=1}^\infty d_n(L_\perp) \left( \frac{\as}{4\pi} \right)^n$,
$L_\perp=\ln\frac{\m^2 b^2}{4e^{-2\g_E}}$,
the cusp anomalous dimension 
$\G_{\rm cusp}=\sum_{n=1}^\infty \G_{n-1}\left( \frac{\as}{4\pi} \right)^n$ 
and the QCD $\b$-function
$\b=\sum_{n=1}^\infty \b_{n-1}\left( \frac{\as}{4\pi} \right)^n$, 
one gets a recursive differential equation for $d_n$.
Solving this equation and performing the all order resummation one gets 
\begin{align}\label{eq:resummedD}
D^R(b;\m_i) &=
-\frac{\Gamma_0}{2\beta_0}\ln(1-X)
+ \frac{1}{2}\le(\frac{a}{1-X}\ri) \le[
- \frac{\beta_1\Gamma_0}{\beta_0^2} (X+\ln(1-X))
+\frac{\Gamma_1}{\beta_0} X\ri]
\nn\\
&+ \frac{1}{2}
\le(\frac{a}{1-X}\ri)^2\le[
2d_2(0)
+\frac{\Gamma_2}{2\beta_0}(X (2-X
))
+\frac{\beta_1\Gamma_1}{2 \beta_0^2} \le( X (X-2)-2 \ln (1-X)\ri)
\ri.
\nn\\
&\le.
+\frac{\beta_2\Gamma_0}{2\beta_0^2} X^2 
+\frac{\beta_1^2\Gamma_0}{2 \beta_0^3} (\ln^2(1-X)-X^2)
\ri]
 +...
\,,
\end{align}
where $a=\as(\m_i)/(4\pi)$ and $X=a\b_0L_\perp$.
The coefficients of $[a/(1-X)]^n$ in the equation above are obtained for $X<1$, however this does not mean that the convergence interval of $D^R$ is the same.
Below we discuss the radius of convergence of $D^R$ in impact parameter space.
For this result we have also used the known first two coefficients of the $D$ term, which can be extracted from known perturbative calculations of the Drell-Yan cross section,
\begin{align}
d_1(0) &= 0
\,,
\nn\\
d_2(0) &=
C_F C_A \left(\frac{404}{27}-14\z_3\right)
- \left(\frac{112}{27}\right)C_F T_F n_f
\,.
\end{align}

The evolution using $D^R$ in Eq.~(\ref{eq:resummedD}) and the usual CSS approach, before the introduction of any non-perturbative model or parameter, resum the same kind of logarithms.
In order to show this, we choose $\mu_i=Q_i$ and $\mu_f=Q_f$.
In CSS approach the $D$ term is resummed using its RG-evolution in Eq.~(\ref{eq:drg}),
\begin{align}\label{eq:devolved0}
D\le(b;Q_i\ri) &=
D\le(b;\m_b\ri)
+ \int_{\m_b}^{Q_i}\frac{d\bar\m}{\bar\m} \G_{\rm cusp}
\,,
\end{align}
choosing $\mu_b=2e^{-\gamma_E}/b$ to cancel all the large $L_\perp$ logarithms.
At lowest order in perturbation theory one gets
\begin{align}
\label{eq:CSSD1}
D(b; Q_i)&=-\frac{\Gamma_0}{ 2 \beta_0}\ln\frac{\alpha_s(Q_i)}{\alpha_s(\mu_b)}\ ,
\end{align}
which expanding $\alpha_s(\mu_b)$ in terms of $\alpha_s(Q_i)$  at the proper perturbative order, $\alpha_s(\mu_b)=\alpha_s(Q_i)/(1-X)$, can be written as
\begin{align}
\label{eq:CSSD2}
D(b;Q_i)&=-\frac{\Gamma_0}{ 2 \beta_0}\ln (1-X)
\,.
\end{align}
This result is the same as the first term of the r.h.s of Eq.~(\ref{eq:resummedD}).
One can also show this equality at higher orders.
Thus, we conclude that $D^R$, given in Eq.~(\ref{eq:resummedD}), coincides with the CSS approach, given in Eq.~(\ref{eq:devolved0}), when all terms in CSS approach are resummed to its appropriate order.

However in practical implementation of CSS method one usually uses formulas like Eq.~(\ref{eq:CSSD1}), and so $\a_s(\mu_b)$ is affected  by the decoupling thresholds of charm and bottom quarks, while in Eq.~(\ref{eq:CSSD2}) these thresholds marginally affect the final result.
In order to go from Eq.~(\ref{eq:CSSD1}) to Eq.~(\ref{eq:CSSD2}), no higher order contributions to the running of the coupling have to be included, and the number of flavors for the running of $\a_s(Q)$ and $\a_s(\mu_b)$ has to be the same.
In~\cite{Echevarria:2012pw} we checked that the solution provided by the $D^R$ is less sensitive to the thresholds, while the direct use of Eq.~(\ref{eq:CSSD1}) leads to undesired divergent behavior for relatively low values of the impact parameter.

\begin{figure}[t]
\begin{center}
\includegraphics[width=0.45\textwidth]{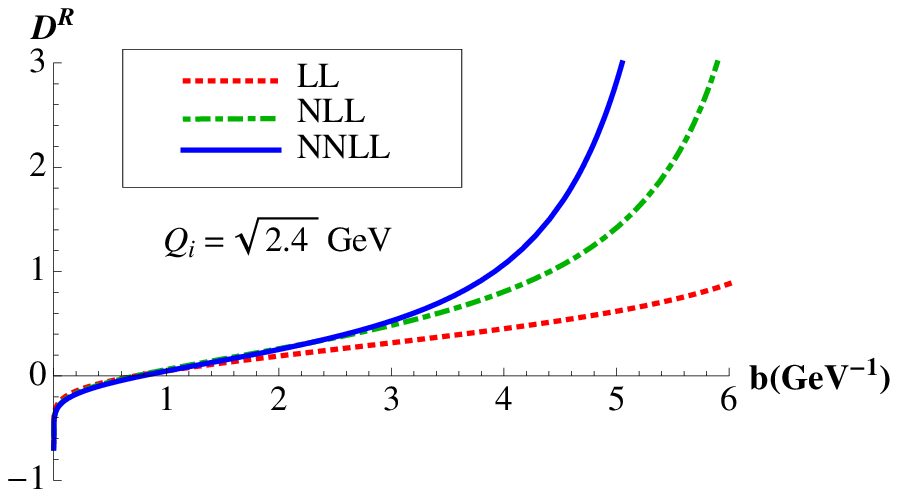}
\quad\quad\quad
\includegraphics[width=0.45\textwidth]{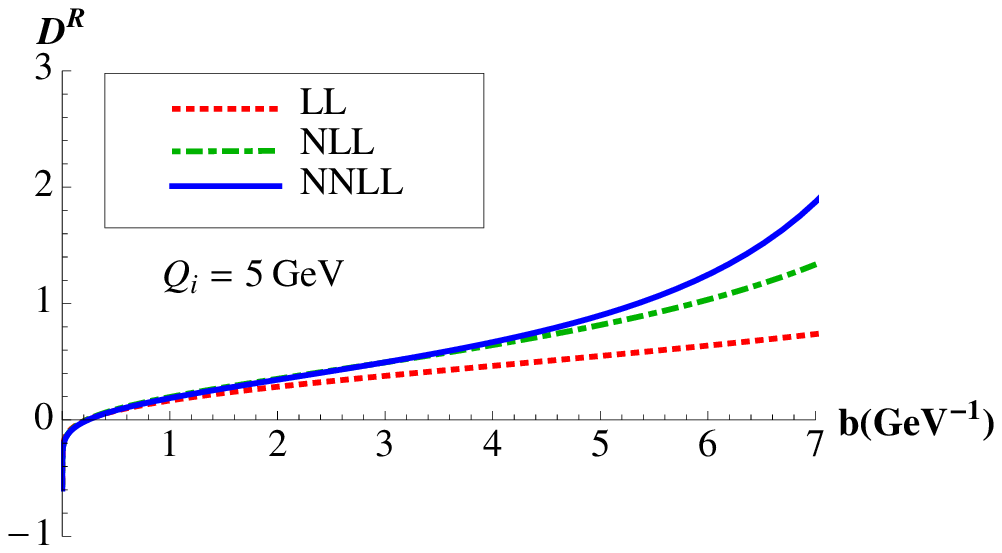}
\\
\hspace{0cm}(a)\hspace{6cm}(b)\hspace{6cm}
\end{center}
\caption{\it
Resummed D at $Q_i=\sqrt{2.4}~{\rm GeV}$ with $n_f=4$ (a) and $Q_i=5~{\rm GeV}$ with $n_f=5$ (b).
}
\label{fig:resummedD}
\end{figure}

In~\cite{Echevarria:2012pw} we discussed as well the separation of perturbative and non-perturbative contributions to the evolution of TMDs. 
Using the $D^R$, we suggest that the evolution kernel can be written as 
\begin{align}\label{eq:Rcnp}
&\tilde R(b;Q_i,\m_i,Q_f,\m_f) =
\nn\\
&
\exp\le\{
\int_{\m_i}^{\m_f} \frac{d\bar\m}{\bar\m} \g_F\le(\as(\bar\m),\ln\frac{Q_f^2}{\bar\m^2} \ri)
\ri\}
\le( \frac{Q_f^2}{Q_i^2} \ri)^
{-\le[D^R\le(b;\m_i\ri)\theta(b_c-b)+D^{NP}(b;\m_i)\theta(b-b_c)\ri]}
\,,
\end{align}
where $b_c$ is some cutoff up to which $D^R$ converges and $D^{NP}$ a non-perturbative input for $b>b_c$. 
The radius of convergence of $D^R$ depends on the initial scale $Q_i$, as can be inferred from Fig.~\ref{fig:resummedD}.
We can thus trust $D^R$ up to $b_c\sim 4~{\rm GeV}^{-1}$ for $\m_i=\sqrt{2.4}~{\rm GeV}$ and $b_c\sim 6~{\rm GeV}^{-1}$ for $\m_i=5~{\rm GeV}$.
The advantage of this implementation of the kernel is that now the perturbative piece of the evolution is treated in a completely perturbative  way, independent from the non-perturbative input or any extra parameters, which renders the results more predictive.

\begin{figure}[t]
\begin{center}
\includegraphics[width=0.45\textwidth]{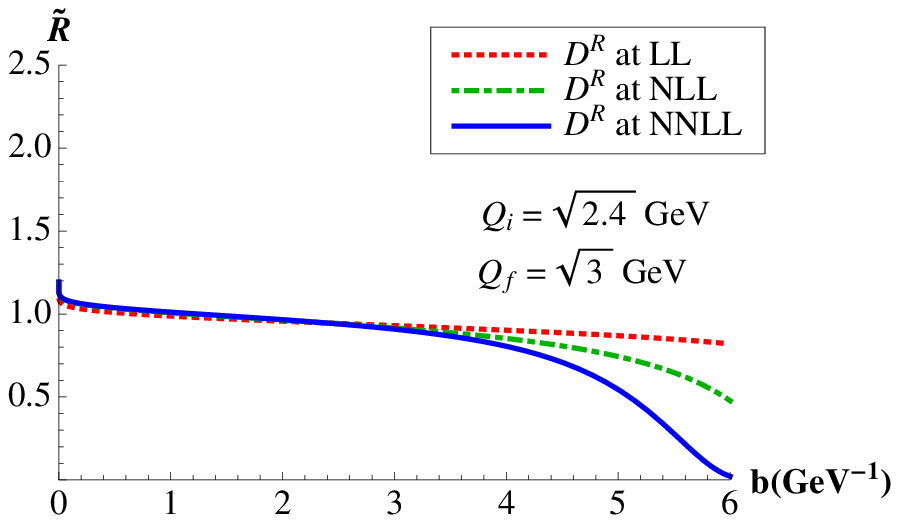}
\quad\quad\quad
\includegraphics[width=0.45\textwidth]{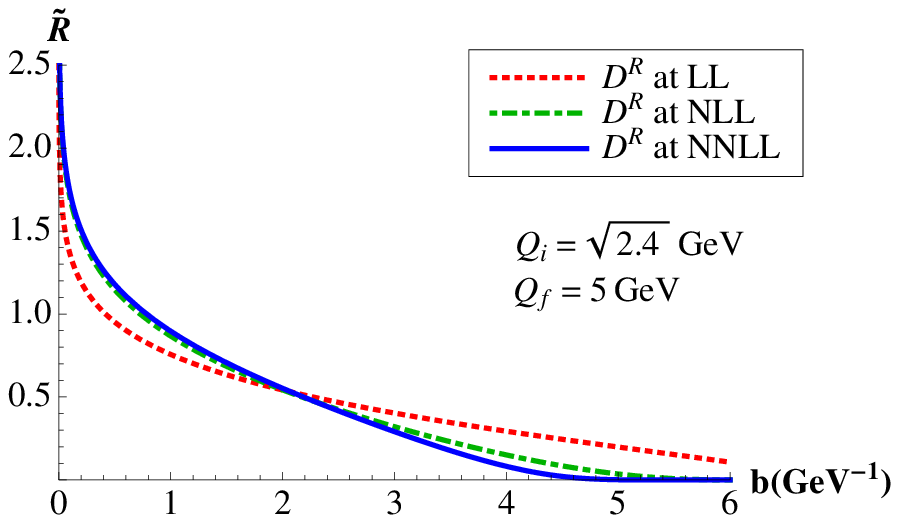}
\\
\hspace{0cm}(a)\hspace{9cm}(b)\hspace{6cm}
\\
\includegraphics[width=0.45\textwidth]{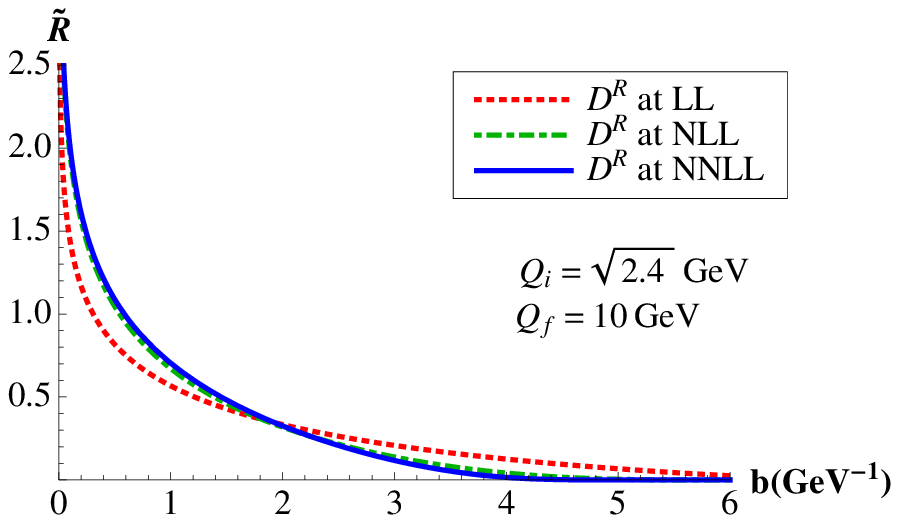}
\quad\quad\quad
\includegraphics[width=0.45\textwidth]{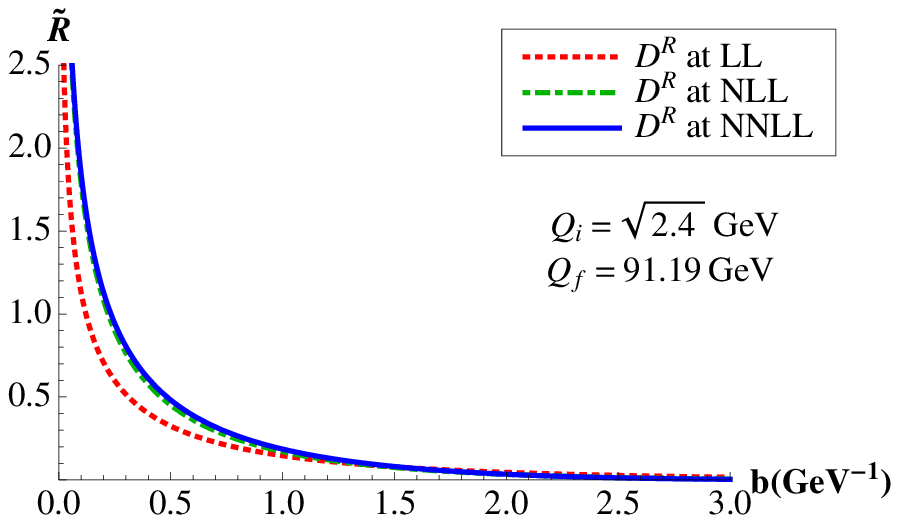}
\\
\hspace{0cm}(c)\hspace{9cm}(d)\hspace{6cm}
\end{center}
\caption{\it
Evolution kernel from $Q_i=\sqrt{2.4}~{\rm GeV}$ up to $Q_f=\{\sqrt{3}\,,5\,,10,91.19\}~{\rm GeV}$.
}
\label{fig:kernelDR}
\end{figure}

In~\cite{Echevarria:2012js} we discuss under which kinematical conditions the effect of $D^{NP}$ in Eq.~(\ref{eq:Rcnp}) can be neglected, or in other words, when it is enough to know the kernel in the perturbative region of the impact parameter.
In this case we could apply a parameter-free expression of the evolution kernel,
\begin{align}\label{eq:Rc}
\tilde R(b;Q_i,\m_i,Q_f,\m_f) &=
\exp\le\{
\int_{\m_i}^{\m_f} \frac{d\bar\m}{\bar\m} \g_F\le(\as(\bar\m),\ln\frac{Q_f^2}{\bar\m^2} \ri)
\ri\}
\le( \frac{Q_f^2}{Q_i^2} \ri)^
{-D^R\le(b;\m_i\ri)} \theta(b_c-b)
\,.
\end{align}
As can be seen from Fig.~\ref{fig:kernelDR}, the larger the final scale $Q_f$ is compared to the initial $Q_i$, the better is the implementation of Eq.~(\ref{eq:Rc}).
In all plots the convergence of the kernel is good up to $b_c\sim 4~{\rm GeV}^{-1}$, which is consistent with Fig.~\ref{fig:resummedD}.
For larger values of $b$ one cannot trust the perturbative expression of the kernel.
However, the larger the final scale is the faster the kernel decreases, thus being not necessary to introduce any non-perturbative model for $b>b_c$.
In other words, our aim is to look for the kinematical setup where the non-perturbative region of the kernel is suppressed, which is the case for $Q_i=\sqrt{2.4}~{\rm GeV}$ and $Q_f\gtrsim 5~{\rm GeV}$.
Using this kernel within the already explained kinematical setup, i.e., as long as $Q_f$ is large enough compared to $Q_i$, we can evolve low energy models for TMDs and extract their parameters by fitting to data.
Thus, all the model dependence is contained in the functional form of the TMDs, while their evolution is parameter-free.

\begin{figure}[t]
\begin{center}
\includegraphics[width=0.45\textwidth]{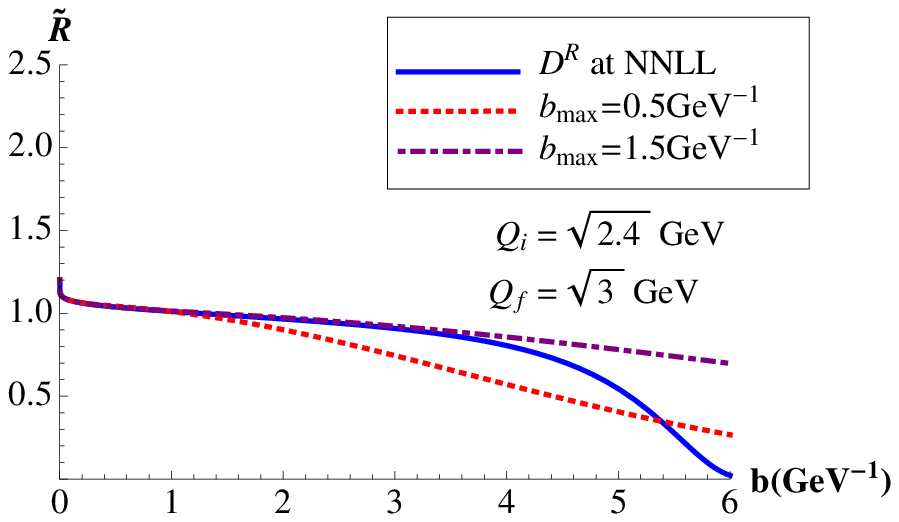}
\quad\quad\quad
\includegraphics[width=0.45\textwidth]{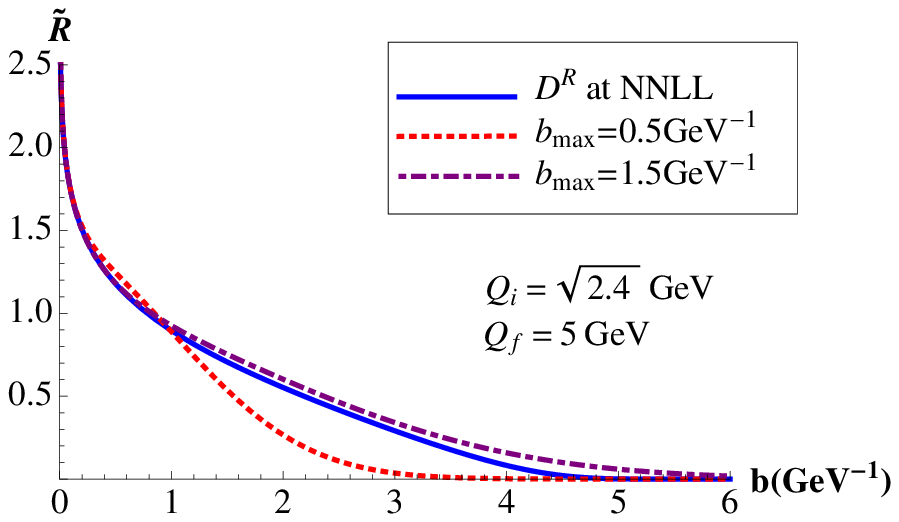}
\\
\hspace{0cm}(a)\hspace{6cm}(b)\hspace{6cm}
\\
\includegraphics[width=0.45\textwidth]{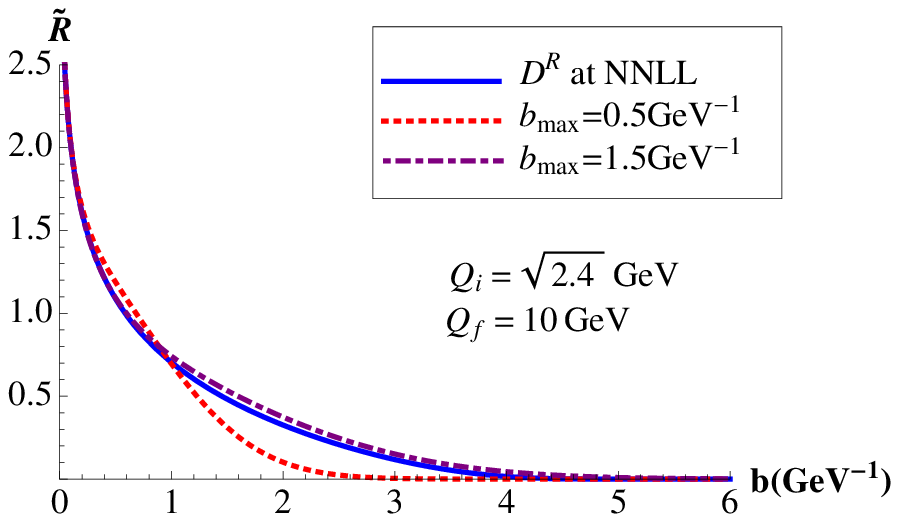}
\quad\quad\quad
\includegraphics[width=0.45\textwidth]{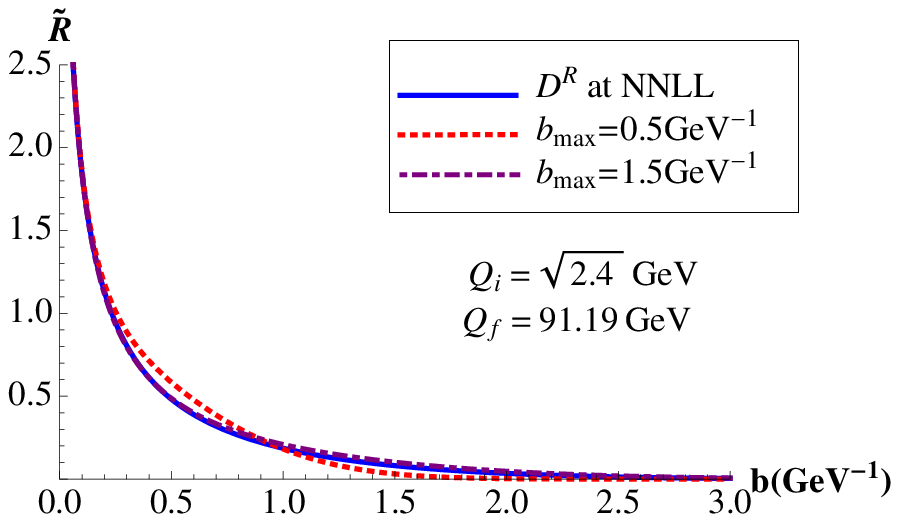}
\\
\hspace{0cm}(c)\hspace{6cm}(d)\hspace{6cm}
\end{center}
\caption{\it
Evolution kernel from $Q_i=\sqrt{2.4}~{\rm GeV}$ up to $Q_f=\{\sqrt{3}\,,5\,,10\,,91.19\}~{\rm GeV}$ using ours and CSS approaches, both at NNLL.
}
\label{fig:kernelDRandCSS}
\end{figure}

In Fig.~\ref{fig:kernelDRandCSS} we compare our approach to the evolution kernel with CSS, both at next-to-next-to leading logarithms (NNLL).
It is clear that Eq.~(\ref{eq:Rc}) can be applied only when the contribution of non-perturbative large $b$ region is negligible, which is the case for large enough $Q_f$.
One can also deduce from all plots that, given the fact that our expression for the evolution kernel is parameter free up to $b_c\sim4~{\rm GeV}^{-1}$, $b_{\rm max}=1.5~{\rm GeV}^{-1}$ gives better results in that region, as was found in~\cite{Konychev:2005iy} by fitting experimental data.
Previous works did not consider $b_{\rm max}$ as a fitting parameter, but rather set it to $0.5~{\rm GeV}^{-1}$ right from the start, fitting just the rest of the parameters of the non-perturbative model.

\begin{figure}[t]
\begin{center}
\includegraphics[width=0.44\textwidth]{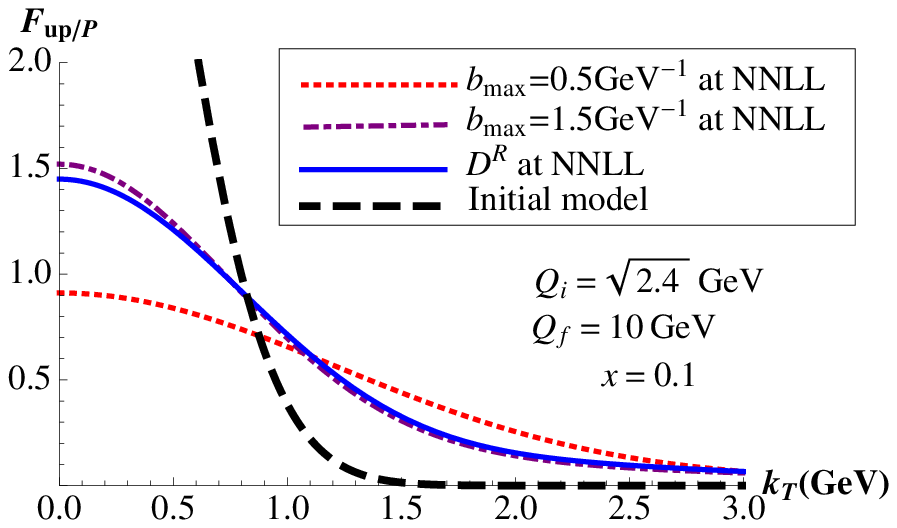}
\\
\hspace{0cm}(a)\hspace{9cm}
\\
\includegraphics[width=0.45\textwidth]{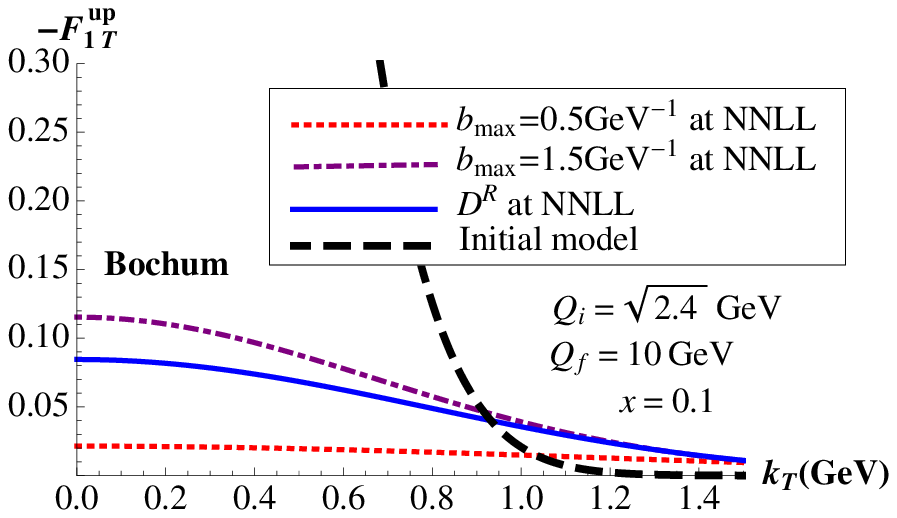}
\quad\quad\quad
\includegraphics[width=0.45\textwidth]{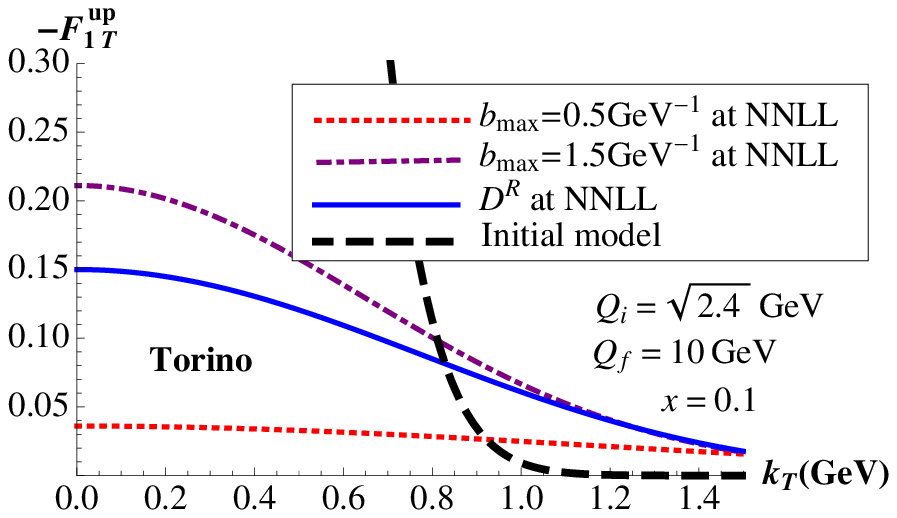}
\\
\hspace{0cm}(b)\hspace{6cm}(c)\hspace{6cm}
\end{center}
\caption{\it
Up quark unpolarized TMDPDF and Sivers function (Bochum and Torino fits) evolved from $Q_i=\sqrt{2.4}~{\rm GeV}$ up to $Q_f=10~{\rm GeV}$ with different approaches to the evolution kernel.
Black line stands for the input Gaussian model and the rest for the evolved TMD either with CSS or our approaches.}
\label{fig:evolvedtmds}
\end{figure}

In order to illustrate the application of the evolution kernel within our formalism and compare it with the CSS approach, we consider existing fits of the unpolarized TMDPDF~\cite{Anselmino:2005nn,Schweitzer:2010tt} and the Sivers function~\cite{Collins:2005ie,Anselmino:2008sga} as inputs.
The unpolarized quark-TMDPDF at low energy is modeled as a Gaussian,
\begin{align}
\tilde{F}_{up/P}(x,b;Q_i) &=
f_{up/P}(x;Q_i)\exp[-\s b_T^2]
\,,
\end{align}
with $\s=0.38/4\,{\rm GeV}^{2}$ for $Q_i=\sqrt{2.4}~{\rm GeV}$~\cite{Schweitzer:2010tt}, and $f_{up/P}$ the up-quark integrated PDF, taken from the MSTW data set~\cite{Martin:2009iq}.
For the Sivers function at low energy we take two different fits, the so-called ``Bochum''~\cite{Collins:2005ie} and ``Torino''~\cite{Anselmino:2008sga} fits in~\cite{Aybat:2011ge}.
The evolved TMDs using our and CSS approaches at NNLL are shown in Fig.~\ref{fig:evolvedtmds}.
The kernel that has been used for $Q_i=\sqrt{2.4}~{\rm GeV}$ and$Q_f=10~{\rm GeV}$ is shown in Fig.~\ref{fig:kernelDRandCSS}c, where one can notice a slight difference between our kernel and the one of CSS with $b_{\rm max}=1.5~{\rm GeV}^{-1}$. 
Since the input TMDPDF is narrower than the input Sivers function, this difference is more noticeable in the latter.
In any case, given the fact that in this kinematical setup the resummed expression for the evolution kernel in Eq.~(\ref{eq:Rc}) is parameter free and convergent, as it is clear from Fig.~\ref{fig:kernelDR}c, the evolved TMDs using our kernel (solid blue lines) should be considered as the most accurate ones.

\section{conclusion}
We have discussed the basic principles behind defining TMDPDFs, which should transcend to other spin-dependent hadronic matrix elements as well.
We also sketched some of the evolution properties of those hadronic matrix elements, where resummation of large logarithms is performed up to NNLL.
One of the basic arguments of this effort is to emphasize that the collinear and soft limits of pQCD contain, individually, rapidity divergences which are unphysical. 
The role of the soft function is to cancel such divergences from the collinear contributions, while the remaining divergences are the perturbative signature of long-distance QCD effects. 

\bibliographystyle{jhep}
\bibliography{references}

\end{document}